\RequirePackage{etoolbox}
\csdef{input@path}{{style/}{graphics/}}
\documentclass[aoas,MSNbibl,nameyear,dvips]{arximspdf}
\usepackage{graphicx}
\usepackage{textcomp}
\doi{10.1214/14-AOAS787} 
\volume{8}
\issue{4}
\pubyear{2014}
\firstpage{2538}
\lastpage{2566}
\docsubty{FLA}

\makeatletter
\newcommand{\lleft}{\left}
\newcommand{\rrvert}{\vert}
\newcommand{\rright}{\right}
\newcommand{\llvert}{\vert}
\newcommand{\overset}{\stackrel}
\newtheorem{theorem}{Theorem}
\newtheorem{corollary}{Corollary}
\newtheorem{lemma}{Lemma}
\newproclaim{assumption}{Assumption}
\makeatother

\begin{document}
\begin{frontmatter}

\title{Nonparametric inference in a stereological model with oriented
cylinders applied to dual phase~steel\thanksref{T1}}
\runtitle{Nonparametric inference for an Oriented Cylinder Model}

\begin{aug}
\author[A]{\fnms{K. S.}~\snm{McGarrity}\corref{}\ead[label=e1]{mcgarrityk@gmail.com}\thanksref{a1,a2}},
\author[B]{\fnms{J.}~\snm{Sietsma}\thanksref{a2}\ead[label=e2]{j.sietsma@tudelft.nl}}
\and
\author[C]{\fnms{G.}~\snm{Jongbloed}\thanksref{a2}\ead[label=e3]{g.jongbloed@tudelft.nl}}
\runauthor{K. S. McGarrity, J. Sietsma and G. Jongbloed}
\affiliation{Materials innovation institute (M2i)\thanksmark{a1} and
Delft University of Technology\thanksmark{a2}}
\address[A]{K. S. McGarrity\\
Materials innovation institute\\
Mekelweg 4 2628CD Delft\\
The Netherlands\\
and\\
Department of Applied Mathematics \\
Delft University of Technology\\
Mekelweg 4 2628CD Delft\\
The Netherlands \\
and\\
Department of Materials Science\\
\quad and Engineering \\
Delft University of Technology\\
Mekelweg 2 2628CD Delft \\
The Netherlands \\
\printead{e1}}
\address[B]{J. Sietsma\\
Department of Materials Science\\
\quad and Engineering \\
Delft University of Technology\\
Mekelweg 2 2628CD Delft \\
The Netherlands\\
\printead{e2}}
\address[C]{G. Jongbloed\\
Department of Applied Mathematics \\
Delft University of Technology\\
Mekelweg 4 2628CD Delft \\
The Netherlands \\
\printead{e3}}
\end{aug}
\thankstext{T1}{Supported under project number M41.10.09330 in the
framework of the Research
Program of the Materials innovation institute M2i
(\surl{www.m2i.nl}).}

\received{\smonth{12} \syear{2013}}
\revised{\smonth{7} \syear{2014}}

%
\begin{abstract}
Oriented circular cylinders in an opaque medium are used to
represent certain microstructural objects in steel. The opaque
medium is sliced parallel to the cylinder axes of symmetry and the
cut-plane contains the observable rectangular profiles of the
cylinders. A one-to-one relation between the joint density of the
squared radius and height of the 3D cylinders and the joint density
of the squared half-width and height of the observable 2D rectangles
is established. We propose a nonparametric estimation procedure to
estimate the distributions and expectations of various quantities of
interest, such as the cylinder radius, height, aspect ratio, surface
area and volume from the observed 2D rectangle widths and heights.
Also, the covariance between the radius and height of a cylinder is
estimated. The asymptotic behavior of these estimators is
established to yield point-wise confidence intervals for the
expectations and point-wise confidence sets for the distributions
of the quantities of interest. Many of these quantities can be
linked to the mechanical properties of the material, and are,
therefore, useful for industry. We illustrate the mathematical model
and estimation procedures using a banded microstructure for which
nearly 90 \textmu m of depth have been observed via serial sectioning.
\end{abstract}

%
\end{frontmatter}

\section{Introduction}
One of the biggest challenges of studying materials like steel is the
inability to see inside of an opaque medium. While there are methods to
obtain three-dimensional (3D) information, they tend to be costly both
in terms of time and resources. Methods like serial sectioning are
destructive to the material and require long periods of time to collect
a reasonable amount of data. Nondestructive methods such as synchotron
radiation are expensive and can only be performed at specialized
laboratories. The discipline of stereology provides many tools to
confront these issues in the sense that there are well established
models that provide means of estimating various 3D quantities based on
(relatively inexpensive) two-dimensional (2D) observations and
measurements; see, for example, \citet
{Mayhew1991,Ohser2000,Russ2000}. A
classical example comes from a study by \citet{Wicksell1925}
where the
size distribution of spherical corpuscles in spleens is estimated based
on measuring the circular cross-sections from slices of the spleens.
Wicksell derived the relationship between the distribution of the
unobservable sphere radii and the distribution of the observable
cross-sectional circle radii. He then used the empirical data and a
histogram estimator to solve his particular problem.

This basic stereological model has been applied in a variety of
disciplines where it is not possible to obtain full 3D measurements of
objects simply by looking at them; this includes biology, geology,
astronomy and materials science: [\citet
{CruzOrive1990,Giumelli1999,Higgins2000,Jeppsson2011,Miyamoto1994,Sahagian1998,Sen2011,Tewari2001}].
Not surprisingly, the method has also gained considerable attention in
the statistics literature. There, the main focus is on computation and
asymptotic behavior of the proposed estimators [\citet
{CruzOrive1985,Mase1995,Sen2011,Silverman1990,vanEs1990}].

In several applications the particles of interest are spheres, or close
enough to be treated as such. However, in many other applications the
particles are not spherical at all, and so it is important to also
consider models with nonspherical particles. The basic model with
spheres has been extended to randomly oriented cylinders, polygons,
spheroids and ellipsoids, and nonregular shapes [\citet
{Andersen2008,Fullman1953,Higgins2000,Giumelli1999,Li1999,Jensen1995,Mehnert1998,Oakeshott1992,Sahagian1998,Spiess2010,Thouless1988}].

\begin{figure}

\includegraphics{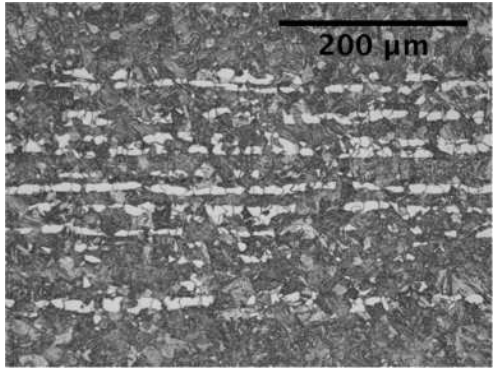}

\caption{Optical image of a banded
steel microstructure.}
\label{figBandedStructure}
\end{figure}

All of this has led to a large body of work from which information of
interest to scientists, engineers and industry can be drawn. The tools
that have been created are powerful in their versatility. They can be
applied to real materials, to models and simulations. They can also be
studied from a theoretical point of view. The specific motivation for
this current work comes from banded steel microstructures, like the one
shown in Figure~\ref{figBandedStructure}. The industry is interested in
this particular material because it has anisotropic properties, high
susceptibility to cracking and corrosion, and it is more difficult to
machine than nonbanded material. This anisotropy can arise either from
the particular chemistry of the steel or during the rolling phase when
blocks of steel are flattened into sheets and rolled into coils.
Currently, there is no reliable way to prevent or control the banding
under certain necessary processing environments. Being able to
quantitatively describe the sizes of the bands in 3D will greatly aid
industry in assessing the quality of the material and the extent of the
effects the bands have on the material coming off the production line.
Ultimately, this will also aid in understanding and controlling the
process that leads to band formation, thereby making it possible to
eliminate them from the material when they are undesirable.

%

In this paper, we propose a simple model in which we use randomly sized,
oriented cylinders to represent the microstructural bands. Following the
example set forth by \citet{Wicksell1925} when he considered spherical
corpuscles observed in spleens, we will consider the marginal
distributions of the radius and height of the cylinders. While most
stereological models assume that nonspherical objects are randomly
oriented, in this case, it is clear that this assumption is not
appropriate. Therefore, by imposing the orientation constraints, we can
explore other properties of the cylinders, such as the volume, surface
area and aspect ratio. These quantities are important to estimate
because they are linked to the mechanical properties of the material.
For example, the surface area can be linked to the interface area
between two phases, which determines properties like strength and
resistance to corrosion or cracking.

In this work, we propose two nonparametric estimators for estimating
the distributions of the 3D cylinder quantities of interest from the 2D
rectangle observations. One estimator enforces a monotonicity
constraint, inspired by the work of \citet{Groeneboom1995}, the other
does not. An empirical estimator is used to estimate the expectations
of the 3D quantities of interest from the 2D observations. The rates of
convergence and asymptotic distributions for all of these estimators are
derived, which provide means of estimating the point-wise confidence
intervals for the expectations and point-wise confidence sets for the
distributions when the model is applied to the steel microstructures.
While a parametric estimator could perform better than the nonparametric
estimators we propose here, not enough is known about the bands within
steel microstructures to assume any particular distribution for the
radius and height of the cylinders. Therefore, the first step toward
understanding this distribution is to study it nonparametrically and so
this work focuses on the empirical and isotonic estimators for
understanding the material.

This paper is organized as follows. The cylinder model is introduced in
Section~\ref{secCylMod}. The nonparametric estimation procedures is
described in Section~\ref{secEstimation} and the asymptotic
distributions and rates of convergence of the two different estimators
are derived in Sections~\ref{secPlug-In} and \ref{secIsotonic}. A
simulation for validation of the model is presented in Section~\ref
{secSimulation} and, finally, in Section~\ref{secReal} the model is
applied to the banded microstructure.

\section{Cylinder model}\label{secCylMod}
%
\begin{figure}

\includegraphics{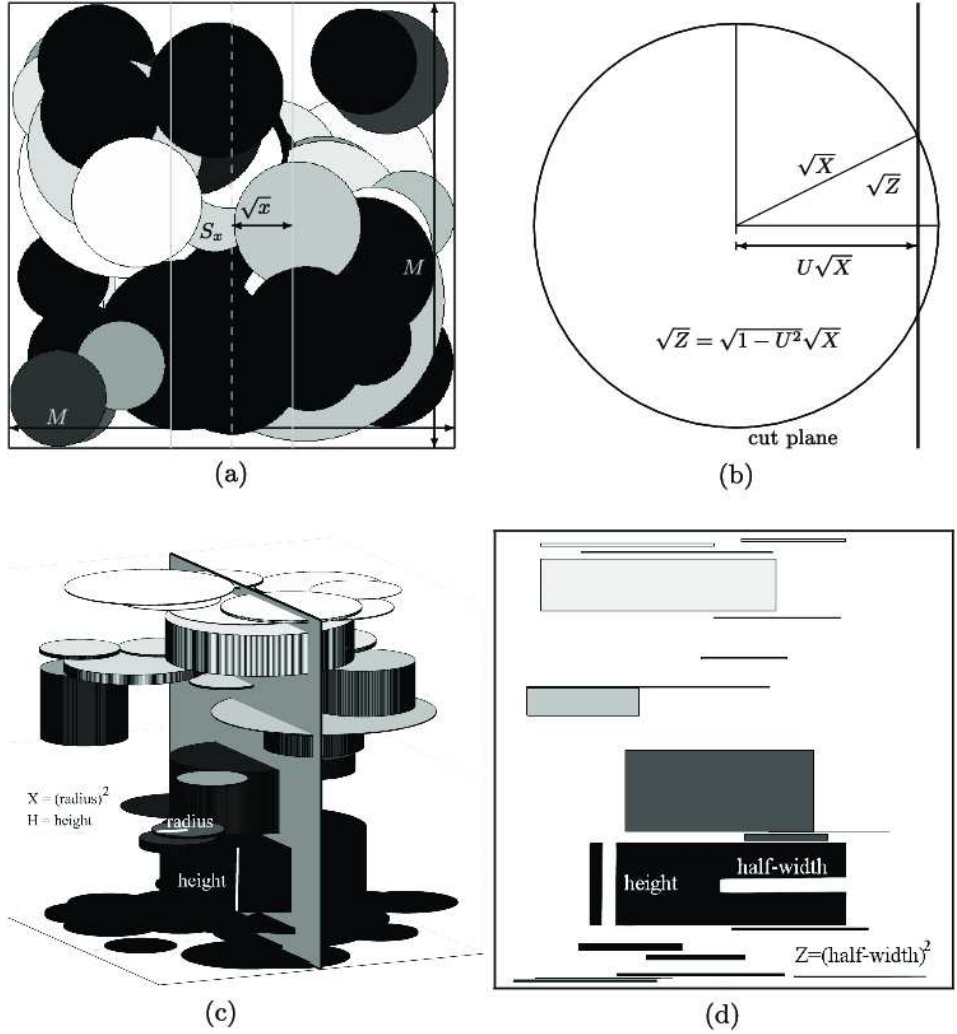}

\caption{Visualization of the cylinder model.
\textup{(a)}~Top view of cylinders in an $M \times M \times M$ box with
a cut plane (dashed line) and slab $S_x$ (solid lines)
into which cylinder centers should fall to be cut by the
plane.
\textup{(b)}~Schematic view, $\sqrt{X}$ is the cylinder radius,
$\sqrt{Z}$ is the rectangle half-width, $U$ is a uniform
random variable.
\textup{(c)}~View of cut plane through the box.
\textup{(d)}~Observations on the cut plane.}\label{figCylinderModel}\label{figSchematic}\label{figBoxPlane}\label{figCutPlane}\label{figTopView}
\end{figure}

To represent the bands shown in Figure~\ref{figBandedStructure}, the
following model is proposed (see Figure~\ref{figCylinderModel}).
Cylinders are generated with a joint density $f$ for the squared
radius $X$ [the choice to look at the \textit{squared} radius is
inspired by \citet{Hall1988}] and height $H$. The centers of these
cylinders are 
cylinders are placed such that their axes of symmetry all have the same
orientation, as in Figure~\ref{figBoxPlane}(c). A cylinder with radius
$\sqrt{x}$ will be intersected by the plane if and only if its center
falls within slab $S_x$ as shown in Figure~\ref{figTopView}(a). This leads
to biased observations on the cut plane since cylinders with larger
radii have a higher probability of being intersected. More specifically,
the joint cumulative distribution function (CDF) of $(X,H)$, given that
the plane intersects the cylinder, can be written as
\begin{eqnarray*}
&& P(X\leq x,H\leq h | \mbox{cylinder hits plane})
\\
&&\qquad = \frac{P(X\leq
x,H\leq h \mbox{ and cylinder hits plane})} {
P(\mbox{cylinder hits plane})}
\\
&&\qquad =\frac{\int_{y=0}^x\int_{m=0}^h \sqrt{y} f(y,m) \,dm \,dy} {
\int_{y=0}^\infty\int_{m=0}^\infty\sqrt{y} f(y,m) \,dm \,dy}
\\
&&\qquad =\frac
{1}{m_F^+} \int_{y=0}^x
\int_{m=0}^h \sqrt{y} f(y,m) \,dm \,dy.
\end{eqnarray*}
Here, since the probability that the cylinder is cut is proportional to
the radius, the density function $f$ is weighted by the ratio of the
radius of the cylinder, $\sqrt{x}$, to the expected radius,
$E_f [\sqrt{X} ] \equiv m_F^+$, which we assume to be finite
(see Assumption \ref{assumpfiniteZ^-12}). Since the centers of the
circles are uniformly distributed throughout the medium, the distance
from the center of a cylinder that has been cut to the intersecting
plane is a uniform random variable, as shown in Figure~\ref
{figSchematic}(b). This is analogous to the relationship between the
circle radii and sphere radii in the method set forth by
\citet{Wicksell1925}. Once a cylinder has been cut, the observable
portion is seen as a rectangle on the cut plane, as shown in
Figure~\ref{figCutPlane}(d).

The rectangles have observable squared half-widths, $z$, and heights,
$h$, that have a joint density $g$. Since the cylinders are all cut
parallel to their axis, all of the height information for the cut
cylinders is preserved and directly observable on the cut-plane. (This
shows that the distribution of the cylinder centers along the
direction of the heights does not require the uniform random
assumption.) The half-widths of the observed rectangles are related to
the cylinder radii through the relationship displayed in Figure~\ref
{figSchematic}(b). From these 2D observations, one can estimate the
3D distribution where the relationship between $g$ and $f$ can be
obtained using a variant of the well-known formula relating the density
of the rectangle half-width (and height) to the distance of cylinder
center to the cut plane and the density of the cylinder radius (and
height):
%
\begin{eqnarray}\label{eqInversef2g}
g(z,h) &=& \frac{\int_{x=z}^\infty(x-z)^{-{1}/2} f(x,h) \,dx} {
2 \int_{x=0}^\infty\sqrt{x} f_X(x) \,dx}
\nonumber\\[-8pt]\\[-8pt]\nonumber
&=&  \frac{1}{2 m_F^+} \int_{x=z}^\infty
(x-z)^{-{1}/2} f(x,h) \,dx.
\end{eqnarray}
This relation can be inverted to obtain the joint density for the
cylinder radius and height as a function of the observable rectangle
joint density:
%
\begin{eqnarray} \label{eqInverseg2f}
f(x,h) &=& -\frac{\partial}{\partial x} \frac{\int_{z=x}^\infty
(z-x)^{-{1}/2} g(z,h) \,dz} {
\int_{z=0}^\infty z^{-{1}/2} g_Z(z) \,dz}
\nonumber\\[-8pt]\\[-8pt]\nonumber
&=& -\frac{1}{m_G^-}
\frac{\partial}{\partial x} \int_{z=x}^\infty
(z-x)^{-{1}/2} g(z,h) \,dz,
\end{eqnarray}
where $m_G^- \equiv E[Z^{-1/2}]$ is the expectation of one over the
rectangle half-width and is also assumed to be finite (see Assumption
\ref{assumpfiniteZ^-12}). From this relationship, the distributions of
univariate quantities of interest such as the height $H$, the squared
radius $X$, the aspect ratio $R = \sqrt{X}/H$, the surface area $S =
2\pi(X+\sqrt{X}H)$, and the volume $V = \pi XH$ can be calculated.

The CDF for the observed height takes on the form
%
\begin{equation}
F_H(h) = \int_{t=0}^h
f_H(t) \,dt = \frac{1}{m_G^-} \int_{t=0}^h
\int_{z=0}^\infty z^{-{1}/2}g(z,t) \,dz \,dt.
\label{eqCDFh}
\end{equation}
Note that this CDF still contains the weight associated with the bias
from the radius of the cylinder. This accounts for any dependence that
might exist between the cylinder height and radius. Should such a
dependence exist, the observed rectangle height distribution will also
be biased. See Figure~\ref{figDistH} and Section~\ref{secHeightEst}
for a more detailed discussion of the biasing of the height observations
associated with a dependence of the height and radius.

For each of the other quantities of interest, define
%
\begin{equation}
\label{eqqh;t} \qquad q(h;t) = \cases{ t, &\quad(squared radius $T=X$),
\vspace*{3pt}\cr
(ht)^2, &\quad(aspect ratio $T=\sqrt{X}/H$),
\vspace*{5pt}\cr
\displaystyle \biggl[\sqrt{\frac{h^2}{4}+\frac{t}{2\pi}}-\frac{h}{2} \biggr]^2, &
\quad $\bigl(\mbox{surface area }T=2\pi(X+\sqrt{X}H)\bigr)$,
\vspace*{3pt}\cr
\displaystyle\frac{t}{\pi h}, &
\quad(volume $T=\pi XH$)}
\end{equation}
[see \hyperref[AproofsWn]{Appendix} for a comprehensive review of the
relationships between $X, H,\break Z$ and $q(h;t)$]. These functions are
chosen such that the random variable of interest $T$ is such that $T>t$
if and only if $X>q(H;t)$ for $h,t > 0$. Hence, using
(\ref{eqInverseg2f}),
%
\begin{equation}
1-F_T(t)= \int_{h=0}^{\infty}\int
_{x=q(h;t)}^{\infty}f(x,h) \,dx\,dh = \frac{N(t)}{N(0)},
\label{eqCDFt}
\end{equation}
where $N$ is a bounded and decreasing function that can be rewritten as
%
\begin{equation}
N(t) = N_{q(\cdot;t)}(t) = \int_{h=0}^\infty\int
_{z=q(h;t)}^\infty\bigl(z-q(h;t)\bigr)^{-{1}/2}
g(z,h) \,dz \,dh. \label{eqNt}
\end{equation}

Note that (\ref{eqNt}) allows for expression of the CDF of the
unobservable 3D cylinder properties in terms of a function $N$ involving
only the joint density $g$ of the observable pair $(Z,H)$. This suggests
natural ways to estimate the CDFs of these quantities, as will be
discussed in Section~\ref{secEstimation}. Also note that under
Assumption \ref{assumpfiniteZ^-12},
%
\begin{equation}
N(t) \leq N(0) = E_g \bigl[Z^{-{1}/2} \bigr] < \infty.
\label{eqN0}
\end{equation}

Along with the distribution functions, it is useful to estimate the
expectations of the quantities of interest. It is especially important
to be able to express these 3D quantities entirely as functions of the
density $g$ of the observable variables $(Z,H)$. This can be done using
equation (\ref{eqInversef2g}) with $\alpha,\beta>-1$ (given that the
moments exist),
%
\begin{eqnarray}\label{eqMoments}
E_g \bigl[Z^\alpha H^\beta\bigr] &=& \int
_{h=0}^\infty\int_{z=0}^\infty
z^\alpha h^\beta g(z,h) \,dz \,dh
\nonumber\\[-8pt]\\[-8pt]\nonumber
&=& \frac{\sqrt{\pi} \Gamma(\alpha+1)} {
2 m_F^+ \Gamma(\alpha+{3}/2)}
E_f \bigl[X^{\alpha+{1}/2} H^\beta\bigr],
\end{eqnarray}
where $m_F^+$ is the same as that given in (\ref{eqInversef2g})
and $\Gamma$ is the Gamma function.

From these cross-moments, another important quantity of interest can be
calculated: the covariance between the radii and heights of the
cylinders. From the moments given in equation (\ref{eqMoments}), the
following expression is obtained for the covariance between the
unobservable radius $\sqrt{X}$ and height $H$ in terms of the observable
rectangle half-width $\sqrt{Z}$ and height $H$:
%
\begin{eqnarray}\label{eqcovariance}
\operatorname{Cov}_f(\sqrt{X},H) &=& \sigma_{\sqrt{X}H}
= E_f [\sqrt{X}H ] - E_f [\sqrt{X}
]E_f[H]
\nonumber\\[-8pt]\\[-8pt]\nonumber
&=& \frac{(\pi/2)E_g[H]} {
E_g [Z^{-{1}/2} ]} - \frac{\pi/2} {
E_g [Z^{-{1}/2} ]} \frac{E_g [Z^{-{1}/2}H ]} {
E_g [Z^{-{1}/2} ]}.
\end{eqnarray}
The stated quantities of interest associated with the density $f$ are
now expressed in terms of the density $g$ of the observable quantities.
The next section will describe empirical and isotonic estimation
procedures that can be used to estimate the unknown distributions and
covariance.

\section{Nonparametric estimation}\label{secEstimation}
The main statistical problem to solve is to estimate the quantities
defined in terms of the joint density $f$, as introduced in
Section~\ref{secCylMod}, based on the observed data from the joint
density $g$.
A natural estimator to begin with in this case is the empirical
or plug-in estimator.

Plugging the empirical distribution of the observed data pairs
$(Z_i,H_i)$ ($1\le i\le n$) into relations (\ref{eqCDFh}) and
(\ref{eqNt}) yields
%
\begin{equation}
\widehat{F}_{H,n}(h) = \frac{\sum_{i=1}^n Z_i^{-{1}/2}1_{[H_i<h]}} {
\sum_{i=1}^n Z_i^{-{1}/2}} \label{eqestFh}
\end{equation}
as an estimator for the CDF of the heights and
%
\begin{equation}
{N}_n(t) = \widetilde{N}_{n,q(\cdot;t)}(t) =
\frac{1}{n}\sum_{i=1}^n
\bigl(Z_i-q(H_i;t)\bigr)^{-{1}/2} 1_{[Z_i>q(H_i;t)]}
\label{eqestN}
\end{equation}
as estimators for the various choices of $N$ dependent on $q(h;t)$.
These estimators of $N$ can be plugged into (\ref{eqCDFt}) to obtain
the estimators for the CDFs of the various quantities of interest.

The expectations of interest in equation (\ref{eqMoments}) can be estimated
by the empirical mean:
%
\begin{equation}
\label{eqMomentsEst} \widehat{E} \bigl[Z^\alpha H^\beta\bigr] =
\frac{1}{n} \sum_{i=1}^n
Z_i^\alpha H_i^\beta.
\end{equation}
In this way, the covariance between $\sqrt{X}$ and $H$ can be estimated
by
%
\begin{eqnarray}
\hat{\sigma}_{n,\sqrt{X}H} &=& \frac{(\pi/2)\sum_{i=1}^n H_i} {
\sum_{i=1}^n Z_i^{-{1}/2}} - \frac{\pi/2} {
n^{-1}\sum_{i=1}^n Z_i^{-{1}/2}}
\frac{\sum_{i=1}^n H_i Z_i^{-{1}/2}} {
\sum_{i=1}^n Z_i^{-{1}/2}}. \label{eqestcov}
\end{eqnarray}

\begin{figure}

\includegraphics{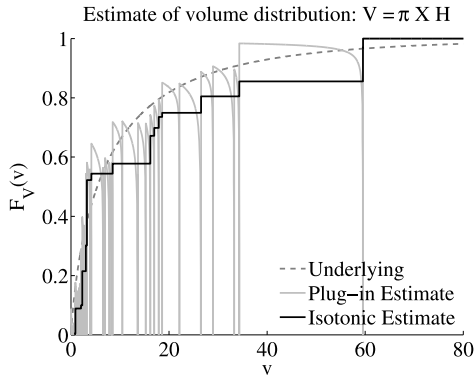}

\caption{The estimates for the underlying
distribution of the volume (given by the simulation in
Section~\protect\ref{secSimulation}) for $n=50$ cylinders. The
underlying distribution is given by the dashed grey line,
the empirical plug-in estimate is given by the solid
light grey line, and the isotonic estimate is given by the
solid black line.}\label{figVol}
\end{figure}

The empirical plug-in estimator works well for estimating the covariance
and yields a monotonic function for the estimate of the distribution
function of the height. This is not true, however, for
$\widetilde{N}_n$. This estimator for $N$, which in view of
(\ref{eqCDFt}) is nonincreasing,\vadjust{\goodbreak} is a nonmonotonic function; it
even has poles due to the vanishing denominator when $q(H_i;t)=Z_i$.
See, for example, Figure~\ref{figVol}. Therefore, inspired by the
approach of \citet{Groeneboom1995}, we introduce an isotonic estimator,
which enforces monotonicity, to obtain estimates for $N$ and,
consequently, the underlying distribution functions of $X$, $R$, $S$ and
$V$.

%

Briefly, the isotonic estimator is the (nonincreasing) function
$\widehat{N}_n$ that minimizes
%
\begin{equation}
N\mapsto\int_0^\infty N(y)^2 \,dy - 2
\int_0^\infty\widetilde{N}_n(y) N(y)
\,dy \label{eqminimize}
\end{equation}
over\vspace*{1pt} all nonincreasing functions on $[0,\infty)$. It is tempting to
``complete the square'' and choose to minimize the function
$\int(N(y)-\widetilde{N}_n(y))^2 \,dy$ instead of (\ref{eqminimize}),
which should lead to the same solution since the added constant,
$\int_0^\infty\widetilde{N}_n(y)^2 \,dy$, does not depend on $N$.
However, $\widetilde{N}_n$ is not square integrable, and so this added
constant is infinite, making this problem ill defined. Therefore, we
stick to minimizing (\ref{eqminimize}).

To solve the minimization problem (continuous isotonic regression), we
use Lemma 2 from \citet{Anevski2009} [see also \citet
{GroJo2010}], where a
characterization is given for the solution of our minimization problem.
We begin by integrating the empirical estimator in (\ref{eqestN}) with
respect to $t$, yielding
%
\begin{eqnarray}\label{eqUnt}
U_n(t) &=& \int_{u=0}^t
\widetilde{N}_n(u) \,du
\nonumber\\[-8pt]\\[-8pt]\nonumber
&=& \int_{u=0}^t
\frac{1}{n} \sum_{i=1}^n
\bigl(Z_i-q(H_i;u)\bigr)^{-{1}/2} 1_{[Z_i>q(H_i;u)]}
\,du.
\end{eqnarray}
Then, define $U_n^*$ to be the least concave majorant of $U_n$,
enforcing monotonicity of its derivative. Finally, for $t \ge0$,
$\widehat{N}_n(t)=U_n^{*,r}(t)$ is the right-hand derivative of~$U_n^*$
evaluated at $t$.

Sections~\ref{secPlug-In} and \ref{secIsotonic} will consider the
rates of convergence and asymptotic distributions for the plug-in
estimators and the isotonic estimator in turn.
\section{Asymptotic distributions of the plug-in estimators}\label{secPlug-In}
There are a few assumptions on the observed variables that are required
for the derivation of consistency and the various asymptotic
distributions to hold.

\begin{assumption}
\label{assumpfiniteZ^-12}
$0 < E_g [Z^{-{1}/2} ] < \infty$. Equivalently, via
(\ref{eqInversef2g}) and (\ref{eqMoments}),
$0 < E_f [\sqrt{X} ] < \infty$.
\end{assumption}

%
\begin{assumption}
\label{assumpfiniteH^5+}
$E_g[H^{5+\varepsilon}] < \infty$ for some $\varepsilon> 0$.
\end{assumption}

%
\begin{assumption}
\label{assumpfiniteHZ^-12}
$E_g [Z^{-{1}/2} H ] < \infty$.
\end{assumption}

Under Assumptions \ref{assumpfiniteZ^-12}, \ref{assumpfiniteH^5+}
and \ref{assumpfiniteHZ^-12}, the plug-in estimators for the
distribution function of $H$, the quantities $N(t)$ for $X$, $R$, $S$
and $V$ (for fixed $t$), and the covariance in equations
(\ref{eqestFh}), (\ref{eqestN}) and (\ref{eqestcov}), respectively,
are consistent by the law of large numbers. From (\ref{eqInversef2g}),
(\ref{eqInverseg2f}) and (\ref{eqMoments}) it follows that the random
variables $Z^{-1/2}$, $HZ^{-1/2}$ and
$[Z-q(H;t)]^{-1/2} 1_{[Z>q(H;t)]}$ have infinite variances. This
means that the standard (finite variance) central limit theorem cannot
be used to derive relevant asymptotic distributions. The theorem below
states a central limit result for random variables with infinite
variances that will be needed in the sequel.

\begin{theorem}\label{thmCylinderCLT}
Let $Y_i$, for $i=1,2,\ldots,$ be i.i.d. random variables. Denote the
distribution of $Y_i$ by $K$ and define
$\overline{Y_n} = \frac{1}{n}\sum_{i=1}^n Y_i$.
If $E_K[Y_i] < \infty$ and
$P_K(Y_i>c)\sim\frac{\kappa}{c^2}$ as $c\rightarrow\infty$ and
$E_K[Y_i^2 1_{[Y_i\in[0,c)]}]\sim\kappa \ln(c^2)$, where
$\kappa> 0$ is a constant, then
\[
\sqrt{\frac{n}{\ln(n)}} \bigl(\overline{Y_n}-E_K[Y_i]
\bigr) \leadsto\mathcal{N}(0,\kappa).
\]
\end{theorem}

\begin{pf}
We apply Theorem 4 from Chapter~9 of \citet{Chow1988}. To this end,
note that because $P_K(Y_i>c) \sim\frac{\kappa}{c^2}$ and
$E_K[Y_i^2 1_{[Y_i\in[0,c)]} ] \sim\kappa\ln(c^2)$, the
following condition holds:
\begin{eqnarray*}
\lim_{c \rightarrow\infty} \frac{\int_{\llvert y\rrvert >c}\,dK(y)} {
({1}/{c^2})\int_{\llvert y\rrvert <c}y^2 \,dK(y)} &=& \lim_{c\rightarrow
\infty}
\frac{P(Y_i>c)}{({1}/{c^2})E_K[Y_i^2 1_{[Y_i\in[0,c)]}]}
\\
&=& \lim
_{c\rightarrow\infty} \frac{\kappa}{\kappa\ln(c^2)} = 0.
\end{eqnarray*}
Now, choose $c=\sqrt{n \ln(n) \kappa}$ and define
$A_n = \frac{n}{B_n} \int_{\llvert y\rrvert <B_n}y \,dK(y)$ and
$B_n = \sup\{c\dvtx\frac{1}{c^2}\int_{\llvert y\rrvert <c}y^2 \,dK(y)
\geq\frac{1}{n} \}$. This leads to
$B_n \sim c$ and
$A_n\sim\sqrt{\frac{n}{\ln(n) \kappa}}E_K[Y_i]$ for $n\rightarrow\infty$
since $E_K[Y_i]<\infty$. Consequently, the central limit theorem holds
where, for $y \in\mathbb{R}$,
\begin{eqnarray*}
\lim_{n\rightarrow\infty}P \Biggl(\frac{1}{B_n}\sum
_{i=1}^n Y_i - A_n < y
\Biggr) &=& \lim_{n\rightarrow\infty} P \biggl(\sqrt{\frac{n}{\ln(n)
\kappa}} \bigl(
\overline{Y_n} - E_K[Y_i]\bigr) < y \biggr)
\\
&=&
\Phi(y),
\end{eqnarray*}
where $\Phi$ is the CDF of the standard normal distribution.
\end{pf}

\subsection{Asymptotic distributions for the estimators of $N(t)$ and $F(t)$}
Using Theorem \ref{thmCylinderCLT}, we derive the asymptotic
distribution for estimators of $N(t)$ for the various choices of $q$
given in (\ref{eqqh;t}). We begin by defining the density
function of the random variable $Z-q(H;t)$ as
%
\begin{equation}
\tau_q(z) = \tau_{q(\cdot;t)}(z) = \int_{h=0}^\infty
g\bigl(z+q(h;t),h\bigr) \,dh. \label{eqtau}
\end{equation}
%

\begin{assumption}
$\tau_q'$ is continuous and uniformly bounded by some $M<\infty$
in a right neighborhood of $0$.
\label{assumpunifcont}
\end{assumption}

If Assumption \ref{assumpunifcont} holds, then (\ref{eqtau}) has the
important property that for $\delta\downarrow0$,
%
\begin{equation}
\int_{z=0}^\delta\tau_q(z) \,dz =
\delta\tau_q(0) + o(\delta). \label{eqassumpunifcont}
\end{equation}

%
\begin{theorem}
\label{thmEmpNt}
Let $(Z_i,H_i)$ ($i=1,2\ldots$) be an i.i.d. sequence with density
$g$ given in (\ref{eqInversef2g}), $t \ge0$ fixed,\vspace*{1pt} and let $q$ be
any of the choices given by (\ref{eqqh;t}). Furthermore,\vadjust{\goodbreak} let
$\widetilde{N}_n$ be defined as in (\ref{eqestN}) and let
Assumption \ref{assumpfiniteZ^-12} hold and Assumption
\ref{assumpunifcont} be satisfied for $q(\cdot;t)$ and $g$. Then
%
\begin{equation}
\sqrt{\frac{n}{\ln(n)}} \bigl(\widetilde{N}_n(t) - N(t) \bigr)
\leadsto\mathcal{N}\bigl(0,\tau_q(0)\bigr).
\end{equation}
\end{theorem}

\begin{pf}
Define\vspace*{1pt} the i.i.d. sequence $Y_1,Y_2,\ldots$ by
$ Y_i = [Z_i-q(H_i;t) ]^{-{1}/2}\* 1_{[Z_i>q(H_i;t)]}$
for $i=1,2,\ldots$ with distribution function $K_Y$. Note that
$\widetilde{N}_n(t) = n^{-1}\sum_{i=1}^n Y_i$ and
$E[Y_i] = N(t) < \infty$ by Assumption \ref{assumpfiniteZ^-12} and
(\ref{eqN0}). The tail probabilities of $Y_i$ behave like
\begin{eqnarray*}
P(Y_i>y) &=& P \biggl(\frac{1_{[Z_i>q(H_i;t)]}} {
\sqrt{Z_i-q(H_i;t)}}>y \biggr)
\\
&= &P
\biggl(q(H_i;t)<Z_i<\frac{1}{y^2}+q(H_i;t)
\biggr)
\nonumber
\\
&=& \int_{h=0}^\infty\int_{z=q(h;t)}^{{1}/{y^2}+q(h;t)}g(z,h)
\,dz \,dh
\\
&=& \int_{h=0}^\infty\int_{z=0}^{{1}/{y^2}}g
\bigl(z+q(h;t),h\bigr) \,dz \,dh
\\
&=& \int_{z=0}^{{1}/{y^2}} \int_{h=0}^\infty
g\bigl(z+q(h;t),h\bigr) \,dz \,dh
\\
&=& \int_{z=0}^{{1}/{y^2}}
\tau_q(z) \,dz = \frac{1}{y^2} \tau_q(0) + o
\bigl(y^{-2} \bigr).
\end{eqnarray*}
Applying (\ref{eqassumpunifcont}) as $y \rightarrow\infty$,
we see that $\kappa= \tau_q(0)$ in Theorem \ref{thmCylinderCLT}.
The expectation of $Y_i^2$ truncated at
$c = \sqrt{n \ln(n) \kappa}$ is
%
\begin{eqnarray}\label{eqE[Y2]}
E\bigl[Y_i^2 1_{Y_i\in[0,c)}\bigr] &=& \int
_{y=0}^{c} y^2 \,dK_Y(y)
\nonumber\\[-8pt]\\[-8pt]\nonumber
&=&
\int_{y=0}^{c}2y \bigl(K_Y(c)-K_Y(y)
\bigr) \,dy \sim\ln\bigl(c^2\bigr) \tau_q(0).
\end{eqnarray}
This relationship is proven in the supplemental article [\citet{suppA}].
Therefore, from Theorem \ref{thmCylinderCLT} the result follows.
\end{pf}

By Theorem \ref{thmEmpNt}, the asymptotic variances for the
estimators $\widetilde{N}_n(t)$ based on the quantities $q$ for the
squared radius, aspect ratio, surface area and volume, respectively,
are given by
%
\begin{eqnarray}\label{eqEmpVar}
&\displaystyle \int_{h=0}^\infty g(t,h) \,dh =
g_Z(t), \qquad\int_{h=0}^\infty g
\bigl(h^2t^2,h \bigr) \,dh,&
\nonumber\\[-8pt]\\[-8pt]\nonumber
&\displaystyle \int_{h=0}^\infty g \biggl( \biggl[\sqrt{
\frac{h^2}{4} +\frac{t}{2\pi}} -\frac{h}{2} \biggr]^2,h
\biggr) \,dh\quad\mbox{and}\quad \int_{h=0}^\infty g
\biggl(\frac{t}{\pi h},h \biggr) \,dh.&
\end{eqnarray}

Note that for the squared radius, result (\ref{eqEmpVar}) is not new.
Since it is independent of height, this result is the same as the result
stated in Theorem 2 by \citet{Groeneboom1995} for spherical particles
in Wicksell's problem. However, for the other quantities of interest,
which require both the squared radius and the height of the cylinders,
the result is different from what can be obtained by following
Groeneboom and Jongbloed's approach to the Wicksell problem. The
asymptotic distributions of $\widetilde{N}_n(t)$ can be used to obtain
the asymptotic distributions of the corresponding distribution
functions of interest, evaluated at $t$. Note that for all choices of
$q$ in (\ref{eqqh;t}), $\widetilde{N}_n(0) = \frac{1}n \sum_{i=1}^n
Z_i^{-1/2}$ and $N(0) = E_g [Z^{-1/2} ] = m_G^- =
\pi/(2m_F^+)$.

\begin{corollary}
\label{corEmpFt}
Based on the estimators $\widetilde{N}_n(t)$ of Theorem
\ref{thmEmpNt}, define
$\widetilde{F}_n(t)=1-\widetilde{N}_n(t)\slash\widetilde{N}_n(0)$ as
estimator for $F_T$ defined in (\ref{eqCDFt}). Then, under the
conditions of Theorem \ref{thmEmpNt}, we have for
$n\rightarrow\infty$
%
\begin{equation}
\sqrt{\frac{n}{\ln(n)}} \bigl(\widetilde{F}_n(t)-F(t) \bigr)
\leadsto\mathcal{N} \biggl(0, \frac{N(0)^2 \tau_q(0) + N(t)^2 g_Z(0)} {
N(0)^4} \biggr).
\end{equation}
\end{corollary}

The proof follows from Theorem \ref{thmEmpNt} using Slutsky's lemma.

\subsection{Asymptotic distribution for the estimator of the covariance}\label{secCov}
Finding the asymptotic distribution of the covariance estimator is more
complicated than for any single expectation estimator. Therefore, this
asymptotic distribution is considered first and the results are then
applied to the simpler estimators for the various expectations. From
Assumption \ref{assumpfiniteH^5+} the variance of $H$ is finite.
Therefore, the standard central limit theorem for finite variance random
variables holds for the sample mean of the $H_i$'s and we can define an
approximating quantity for the covariance that depends only on the terms
involving $Z^{-1/2}$ [compared to (\ref{eqestcov})]:
\[
\widetilde{\sigma}_{n,\sqrt{X}H} = \frac{(\pi/2) E_g[H_i]} {
n^{-1}\sum_{i=1}^n Z_i^{-{1}/2}} - \frac{\pi/2}{n^{-1}\sum_{i=1}^n
Z_i^{-{1}/2}}
\frac{\sum_{i=1}^n H_i Z_i^{-{1}/2}} {
\sum_{i=1}^n Z_i^{-{1}/2}}.
\]
Note that
$ \delta_n^{-1} (\hat{\sigma}_{n,\sqrt{X}H}
- \widetilde{\sigma}_{n,\sqrt{X}H} ) \overset{P}\rightarrow0$,
where $\delta_n = \sqrt{\frac{\ln(n)}{n}}$. Hence, to derive the
asymptotic distribution of
$ \delta_n^{-1} (\hat{\sigma}_{n,\sqrt{X}H}
- {\sigma}_{n,\sqrt{X}H} )$,
it suffices to derive the asymptotic distribution of
$ \delta_n^{-1} (\widetilde{\sigma}_{n,\sqrt{X}H}
- {\sigma}_{n,\sqrt{X}H} )$.
Considering this distribution, define the function
$\phi\dvtx(0,\infty)^2 \mapsto\mathbb{R}$ as
\[
\phi(u,v) = \frac{\pi}{2} \biggl(\frac{E_g[H]}{u}-\frac{v}{u^2}
\biggr).
\]
Moreover, define
%
\begin{equation}
T_n = \frac{1}{n} \sum_{i=1}^n
\pmatrix{ Z_i^{-{1}/2}
\vspace*{3pt}\cr
H_i Z_i^{-{1}/2}},
\label{eqTndef}
\end{equation}
leading to $\widetilde{\sigma}_{n,\sqrt{X}H} = \phi(T_n)$. In order to pin
down the asymptotic variance of $\widetilde{\sigma}_{n,\sqrt{X}H}$, we need
two more assumptions and the following lemma.

\begin{assumption}
\label{assumpfiniteH^i|Z=0}
$\xi_g^j = \int_{h=0}^\infty h^j g(0,h) \,dh < \infty$ for $j=0,1,2$.
\end{assumption}

%
\begin{assumption}
\label{assumpfinitedgdz}
For some constant $K < \infty$,
$\llvert\frac{\partial}{\partial z}g(z,h)\rrvert\leq K$
for all \mbox{$z,h \geq0$}.
\end{assumption}
%

\begin{lemma}
\label{lembivarCLT}
Let $T_n$ be as defined in (\ref{eqTndef}). Assume that Assumptions
\ref{assumpfiniteZ^-12}, \ref{assumpfiniteH^5+},
\ref{assumpfiniteHZ^-12}, \ref{assumpfiniteH^i|Z=0} and
\ref{assumpfinitedgdz} hold, then
%
\begin{equation}
\delta_n^{-1} \bigl(T_n - E_g[T_n]
\bigr) \leadsto\mathcal{N}(0,\Xi)\qquad\mbox{where }\Xi=
\pmatrix{
\xi_g^0 & \xi_g^1
\vspace*{3pt}\cr
\xi_g^1 & \xi_g^2}
\label{eqCovMatrix}
\end{equation}
and the entries in $\Xi$ can be formulated from (\ref{eqMoments}) to
yield
%
\begin{equation}
\xi_g^j = \int_{h=0}^\infty
h^j g(0,h) \,dh = \frac{E_f [X^{-{1}/2} H^j ]} {
2 E_f [X^{1/2} ]}. \label{eqxigj3D}
\end{equation}
\end{lemma}

The proof of this lemma can be found in the supplemental article
[\citet{suppA}]. We now apply the $\Delta$-method to the quantity
$\phi(T_n)$, which yields
\[
\delta_n^{-1} (\widetilde{\sigma}_{n,\sqrt{X}H} -
\sigma_{\sqrt{X}H} ) = \delta_n^{-1} \bigl(
\phi(T_n) - \phi\bigl(E_g[T_n] \bigr) \bigr)
\leadsto\mathcal{N}\bigl(0,\nu^2\bigr),
\]
where
\[
\nu^2 = \bigl(\nabla\phi\bigl(E_g [T_n ]
\bigr) \bigr)^T \Xi \bigl(\nabla\phi\bigl(E_g
[T_n ] \bigr) \bigr)
\]
and
\[
\nabla\phi(u,v) = \pmatrix{ \displaystyle\frac{\partial}{\partial u}\phi(u,v)
\vspace*{5pt}\cr
\displaystyle \frac{\partial}{\partial v}
\phi(u,v)} = \frac{\pi}{2}\frac{1}{u^3}\pmatrix{ 2v -
E_g[H]u
\vspace*{3pt}\cr
-u}.
\]

This provides $\nu^2$ in terms of the joint densities of the observable
variables:
%
\begin{eqnarray}\label{eqnug^2}
\nu^2 &=& \biggl(\frac{\pi}{2} \biggr)^2
E_g^{-4} \bigl[Z^{-{1}/2}
\bigr]\nonumber
\\
&&{}\times \biggl\{ \xi_g^0 \biggl(4\frac{E_g^2 [Z^{-{1}/2}H ]} {
E_g^2 [Z^{-{1}/2} ]} - 4
\frac{E_g [Z^{-{1}/2}H ]E_g[H]} {
E_g [Z^{-{1}/2} ]} + E_g^2[H] \biggr)
\\
&&\hspace*{89.5pt}{}+ 2
\xi_g^1 \biggl(E_g[H]-\frac{E_g [Z^{-{1}/2}H ]} {
E_g [Z^{-{1}/2} ]}
\biggr) + \xi_g^2 \biggr\}.\nonumber
\end{eqnarray}
Given the cross-moment relationships in (\ref{eqMoments}) and
(\ref{eqxigj3D}), $\nu^2$ can also be expressed in terms of the
underlying joint distribution of the cylinder radii and heights:
%
\begin{eqnarray}\label{eqnuf^2}
\nu^2 &=& \biggl(\frac{\pi}{2}
\biggr)^{-2}
E_f \bigl[X^{1/2} \bigr]\nonumber
\\
&&{}\times \bigl
\{E_f \bigl[X^{-{1}/2} \bigr] \bigl(4E_f^2
\bigl[X^{1/2} \bigr]E^2_f[H] \nonumber
\\
&&\hspace*{69pt}{} -4E_f[H]E_f \bigl[X^{1/2} H \bigr]
E_f \bigl[X^{1/2} \bigr]
+E_f^2
\bigl[X^{1/2} H \bigr] \bigr)
\\
&&\hspace*{16pt} {}+ 2E_f
\bigl[X^{-{1}/2} H \bigr] \bigl(E_f \bigl[X^{1/2} H
\bigr] E_f \bigl[X^{1/2} \bigr]
- E_f[H]
E_f^2 \bigl[X^{1/2} \bigr] \bigr) \nonumber
\\
&&\hspace*{167pt}{} +
E_f \bigl[X^{-{1}/2}H^2 \bigr]
E_f^2 \bigl[X^{1/2} \bigr] \bigr\}.
\nonumber
\end{eqnarray}
This proves the following theorem for the plug-in estimator for
$\sigma_{\sqrt{X}H}$.

\begin{theorem}
\label{thmlimcovar}
Let $\sigma_{\sqrt{X}H}$ and $\hat{\sigma}_{n,\sqrt{X}H}$ be defined as
in (\ref{eqcovariance}) and (\ref{eqestcov}), respectively. Under
the assumptions of Lemma \ref{lembivarCLT}, for $\nu^2$ defined in
(\ref{eqnug^2}) and (\ref{eqnuf^2}),
\[
\sqrt{\frac{n}{\ln n}} (\hat{\sigma}_{n,\sqrt{X}H} -\sigma_{\sqrt{X}H} )
\leadsto\mathcal{N} \bigl(0,\nu^2 \bigr) \qquad\mbox{as } n
\rightarrow\infty.
\]
\end{theorem}

\subsection{Estimating the expectations}
From (\ref{eqMoments}) and (\ref{eqMomentsEst}), it is simple to
verify that the various 3D quantities of interest are given by the 2D
observable quantities with their empirical estimators given in
Table~\ref{tabexpectations}.

%
\begin{table}[b]
\tabcolsep=0pt
\caption{Expectations and empirical
estimates of the 3D quantities of interest given as
functions of the expectations and empirical estimates of
the 2D observable quantities}\label{tabexpectations}
\begin{tabular*}{\tablewidth}{@{\extracolsep{\fill}}@{}lll@{}}
\hline
\textbf{Quantity of interest ($\bolds{T}$)} & \multicolumn{1}{c}{\textbf{Expectation $\bolds{E_f[T]}$}} & \multicolumn{1}{c@{}}{\textbf{Empirical estimator $\bolds{\widehat{E}_f[T]}$}}\\
\hline
Radius: $X^{1/2}$
& $(\pi/2) (E_g [Z^{-{1/2}} ] )^{-1}$
& $(\pi/2) ((1/n)\sum_{i=1}^n
Z_i^{-{1/2}} )^{-1}$
\\[2pt]
Squared radius: $X$
& $ (2E_g [Z^{1/2} ] )
(E_g [Z^{-{1/2}} ] )^{-1}$
& $ (2\sum_{i=1}^n Z_i^{{1/2}} )
(\sum_{i=1}^n Z_i^{-{1/2}} )^{-1}$
\\[2pt]
Height: $H$
& $ (E_g [Z^{-{1/2}} H ] )
(E_g [Z^{-{1/2}} ] )^{-1}$
& $ (\sum_{i=1}^n Z_i^{-{1/2}} H_i )
(\sum_{i=1}^n Z_i^{-{1/2}} )^{-1} $
\\[2pt]
Volume: $\pi XH$
& $ (2\pi E_g [Z^{1/2} H ] )
(E_g [Z^{-{1/2}} ] )^{-1}$
& $ (2\sum_{i=1}^n Z_i^{{1/2}} H_i )
(\sum_{i=1}^n Z_i^{-{1/2}} )^{-1} $
\\[4pt]
Surface area:
& $2\pi[ (2 E_g [Z^{1/2} ] ) (E_g [Z^{-{1/2}} ] )^{-1}$ & $2\pi[ (2\sum_{i=1}^n Z_i^{{1/2}} H_i ) (\sum_{i=1}^n Z_i^{-{1/2}} )^{-1} $
\\[2pt]
\quad $2\pi(X+X^{1/2} H )$ &\quad ${} + (\pi/2) E_g[H]$  &\quad ${} + (\pi/2) (\sum_{i=1}^n H_i )$\\[2pt]
&\quad ${}\times (2E_g [Z^{-{1/2}} ] )^{-1} ] $          & \quad ${}\times  (\sum_{i=1}^n Z_i^{-{1/2}} )^{-1} ] $
\\[4pt]
Aspect ratio:  & $ (\pi E_g [H^{-1} ] ) (E_g [Z^{-{1/2}} ] )^{-1} $
& $\pi (\sum_{i=1}^n H_i^{-1} ) (\sum_{i=1}^n Z_i^{-{1/2}} )^{-1} $\\
\quad $X^{1/2} H^{-1}$\\
\hline
\end{tabular*}
\end{table}

Due to the dependence of the aspect ratio on $H^{-1}$, several more
assumptions are required to continue this analysis. For brevity and
simplicity, the expectation of the aspect ratio will not be considered
any further.

To obtain the asymptotic distributions, Lemma \ref{lembivarCLT} and
the delta method can be used with the following assumption.

\begin{assumption}
$E_g [Z^{1/2} H^j ] < \infty$ and
$E_g [ (Z^{1/2} H^j )^2 ] < \infty$,
where $j=0,1$.
\label{assumpMomZ^12H^j}
\end{assumption}

Under Assumption \ref{assumpMomZ^12H^j}, the expectations can be
treated as constants in the modified function $\phi$, as discussed for
the expectation of the height in the previous section. The
coefficients $s$ and $t$ for linearizing (\ref{eqTndef}) are taken to
be zero where appropriate. Then, the asymptotic variance for the
estimation of the quantities of interest given above is listed in
Table~\ref{tabnu2}.

%
\begin{table}
\tabcolsep=0pt
\caption{Asymptotic variances $\nu_g^2$ from Corollary \protect\ref{corlimExpectations}}\label{tabnu2}
\begin{tabular*}{\tablewidth}{@{\extracolsep{\fill}}@{}ll@{}}
\hline
\textbf{Quantity of interest} & \multicolumn{1}{c@{}}{\textbf{Asymptotic variance $\bolds{\nu_g^2}$}}\\
\hline
Radius & $ (\pi/2 )^2 \xi_g^0 (E_g [Z^{-{1/2}} ] )^{-4} $\\[2pt]
Squared radius & $4 \xi_g^0 (E_g [Z^{1/2} ] )^2 (E_g [Z^{-{1/2}} ] )^{-4} $ \\[4pt]
Height & $ \{\xi_g^0 (E_g [Z^{-{1/2}} H ] )^2 - 2 \xi_g^1 E_g [Z^{-{1/2}} H ] E_g [Z^{-{1/2}} ]$\\[2pt]
       &\quad ${} + \xi_g^2 (E_g [Z^{-{1/2}} ] )^2\} (E_g [Z^{-{1/2}} ] )^{-4} $ \\[4pt]
Volume & $4 \pi^2 \xi_g^0 (E_g [Z^{1/2} H ] )^2 (E_g [Z^{-{1/2}} ] )^{-4} $ \\[2pt]
Surface area & $\xi_g^0 (4\pi E_g [Z^{1/2} ]
+\pi^2 E_g[H] )^2
(E_g [Z^{-{1/2}} ] )^{-4} $\\[2pt]
\hline
\end{tabular*}
\end{table}
This leads to the following corollary to Theorem \ref{thmlimcovar}.

\begin{corollary}
\label{corlimExpectations}
Let $E_f[T]$ and $\widehat{E}_f[T]$ be defined as in Table~\ref
{tabexpectations}, where $T$ is any of the quantities of
interest listed in Table~\ref{tabexpectations}. Under the
assumptions of Lemma~\ref{lembivarCLT} and Assumption
\ref{assumpMomZ^12H^j}, for $\nu_g^2$ as defined in Table~\ref{tabnu2},
\[
\sqrt{\frac{n}{\ln(n)}} \bigl(\widehat{E}_f[T] -
E_f[T] \bigr) \leadsto\mathcal{N} \bigl(0,\nu_g^2
\bigr)\qquad\mbox{as } n \rightarrow\infty.
\]
\end{corollary}

Theorem \ref{thmlimcovar} and Corollary \ref{corlimExpectations}
show that the expectations of the quantities of interest can be
estimated consistently with a rate of $\sqrt{\ln(n)/n}$. These
results can be used to obtain the 95\% confidence intervals for the
unknown expectations being estimated by $\widehat{E}_f[T]$:
%
\begin{equation}
\widehat{E}_f[T] \pm1.96 \nu_g \sqrt{
\frac{\ln(n)}{n}}. \label{eqConfidenceInterval}
\end{equation}
For a discussion on the small sample properties and coverage
probability of the confidence intervals, see Chapter~4 of
\citet{MeThesis} and the supplemental article [\citet{suppA}].
\subsection{Asymptotic distribution for the estimator of the height 
distribution}\label{secHeightEst}

Consider the plug-in estimator for the distribution function of heights,
given in (\ref{eqestFh}). As mentioned before, under Assumption
\ref{assumpfiniteH^5+}, the law of large numbers immediately gives
that $\widehat{F}_{H,n}(h)\overset{P}\rightarrow F_H(h)$ as
$n\rightarrow\infty$. The asymptotic distribution is given in the
theorem below.

\begin{theorem}\label{thmFHh}
Consider $F_H(h)$ and $\widehat{F}_{H,n}(h)$ as given in
(\ref{eqCDFh}) and (\ref{eqestFh}), respectively. Under Assumptions
\ref{assumpfiniteZ^-12} and \ref{assumpfiniteH^i|Z=0},
\[
\sqrt{\frac{n}{\ln n}} \bigl(\widehat{F}_{H,n}(h) - F_H(h)
\bigr) \leadsto\mathcal{N}\bigl(0,\nu^2\bigr),
\]
where $\nu^2 = (m_G^- )^{-2}
(F_H(h) \int_h^{\infty} g(0,y) \,dy
+ (1-F_H(h)) \int_0^h g(0,y) \,dy )$.
\end{theorem}

\begin{pf}
Consider the random vectors
\[
T_n=\frac{1}n\sum_{i=1}^n
\pmatrix{ Z_i^{-{1}/2}
\vspace*{3pt}\cr
Z_i^{-{1}/2}
1_{[H_i<h]}}
\]
with
\[
E [T_n ]=\pmatrix{
m_G^-
\vspace*{3pt}\cr
\displaystyle\int_{z=0}^\infty\int
_{y=0}^h z^{-{1}/2} g(z,y) \,dy \,dz}.
\]
For $T_n$ it is shown in the supplemental article [\citet{suppA}] that
%
\begin{equation}
\label{eqconvTnvec} \sqrt{\frac{n}{\ln n}} \bigl(T_n-E
[T_n ] \bigr) \leadsto{\mathcal N} (0,\Xi),
\end{equation}
where the entries of $\Xi$ are given by
$\xi_{12} = \xi_{21} = \xi_{22} = \int_{y=0}^h g(0,y) \,dy$ and
$\xi_{11} = g_Z(0)$. The result follows by applying the $\Delta$-method
to the function $\phi(u,v) = v/u$ at $T_n$, yielding asymptotic normality
with variance $\nu^2$.
\end{pf}

The estimator for the distribution of the heights given in
(\ref{eqestFh}) accounts for any dependence between the radius and
height of the cylinders. Any correlation that might exist will lead to
the height observations being biased like the rectangle half-width
observations due to the larger cylinders being more likely to be
intersected by the cut plane. However, if the heights are known to be
independent from the cylinder radius, then the biasing in the problem
has no consequences for the distribution of observable heights and we
may simply take the empirical distribution of the observed heights to be
the estimate of the actual distribution:
\[
\widehat{F}_H(h) = \frac{1}{n} \sum
_{i=1}^n 1_{[H_i \leq h]}.
\]
%

%
\begin{figure}

\includegraphics{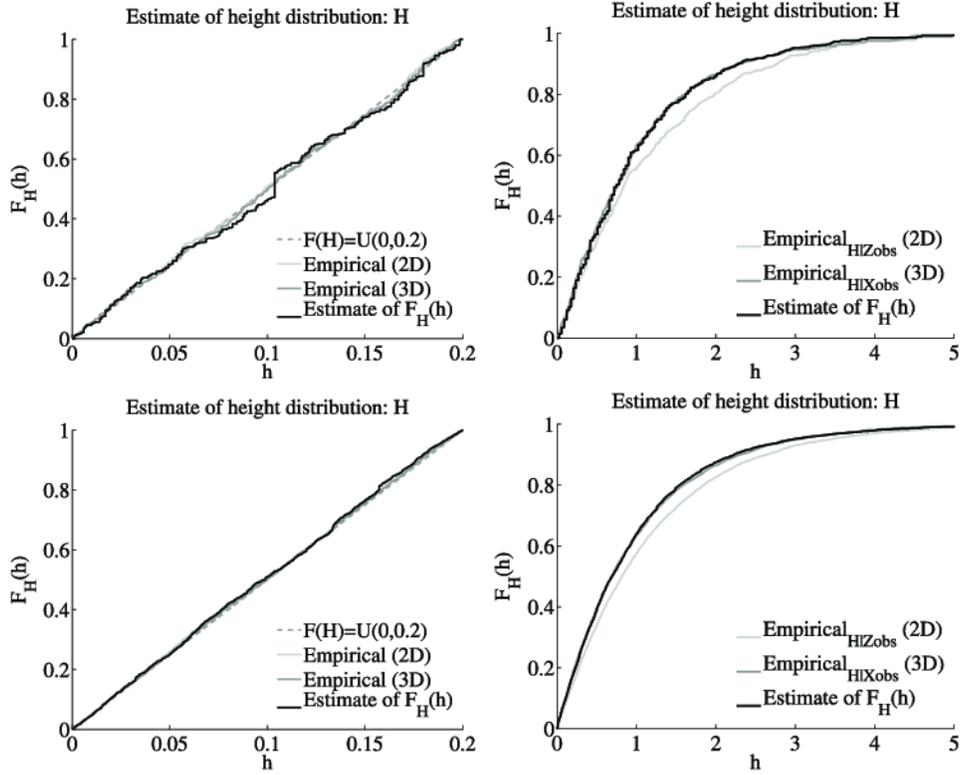}

\caption{The upper and lower figures show the
estimate of $F_H(h)$ for $n=500$ and $n=5000$ cylinders,
respectively. The dark grey lines show the 3D empirical
distributions of the cylinder heights. The light grey lines
show the 2D empirical distributions. The black lines show
the estimates of the 3D distributions as calculated from
(\protect\ref{eqestFh}). The left images are of cylinders whose
height and radii are uncorrelated and the underlying
distribution of the height is shown by the grey dot-dashed
line. The right images are of cylinders whose height and
radii are correlated.}
\label{figDistH}
\end{figure}

\noindent This estimator has a rate of convergence of $1/\sqrt{n}$. Figure~\ref
{figDistH} shows the effect of the rate of convergence for the
estimation procedure. Focusing on the left images, the upper image shows
the 2D (light grey line) and 3D (dark grey line) empirical distributions
for the heights of 500 uncorrelated (radii and cylinder heights),
uniformly distributed cylinders. The bottom image shows the same for
5000 cylinders. The black solid line shows the estimation of the 3D
distribution as calculated from (\ref{eqestFh}). The empirical
distribution is a better choice than (\ref{eqestFh}) in this case
because it has the faster rate of convergence. Contrarily, focusing on
the right images where there is a nonzero correlation between the radii
and heights of the cylinders, the biasing in the 2D distribution
(light grey lines compared to the dark grey line for the 3D empirical
distribution) is clear. In this case, the estimator from
(\ref{eqestFh}) is necessary to accurately estimate the underlying 3D
distribution.

\section{Asymptotic distributions of the isotonic estimators}
\label{secIsotonic}
In this section we study the consistency and asymptotic behavior of the
isotonic estimators, $\widehat{N}_n$, as described in Section~\ref
{secEstimation}. To do so requires one further assumption.

%
\begin{assumption}
\label{assumpfiniteUn}
$\int_0^\infty N(t) \,dt < \infty$.
\end{assumption}

%
\begin{theorem}
\label{thmIsoCLT}
Suppose $t\geq0$ and $F_T$ from (\ref{eqCDFt}) has a density $f$
that is strictly positive and continuous in a neighborhood of $t$ (a
right neighborhood if $t=0$) and that $q(h;t)$ is defined as in
(\ref{eqqh;t}). Further, suppose that Assumptions
\ref{assumpfiniteZ^-12}, \ref{assumpunifcont} and
\ref{assumpfiniteUn} hold. Then,
%
\begin{equation}
\sqrt{\frac{n}{\ln n}} \bigl(\widehat{N}_n(t)-N(t) \bigr)
\leadsto\mathcal{N} \biggl(0,\frac{1}2 \tau_q(0) \biggr)
\end{equation}
as $n \rightarrow\infty$.
\end{theorem}

The proof of this theorem can be found in the supplemental article
[\citet{suppA}]. The striking difference with Theorem \ref
{thmEmpNt} is
the factor $1/2$ in the asymptotic variance. This means that enforcing
monotonicity in the estimator improves on the empirical estimator
because the resulting estimator satisfies the natural monotonicity
constraint. Moreover, it also leads to a more accurate estimator
asymptotically.

Analogous to Corollary \ref{corEmpFt}, we have the following corollary.

\begin{corollary}
\label{corIsoFt}
Suppose that $q(h;t) > 0$ for all $h$ and $t>0$, and that $F_T(t)$ has
a density $f$ which is strictly positive at $t$ and continuous in a
neighborhood of~$t$. Then, under the assumptions of Theorem
\ref{thmIsoCLT},
%
\begin{equation}
\qquad \sqrt{\frac{n}{\ln(n)}} \biggl(1-\frac{\widehat{N}_n(t)}{\widehat
{N}_n(0)}-F_T(t)
\biggr) \leadsto\mathcal{N} \biggl(0,\frac{N(0)^2\tau_q(0) +
N(t)^2g_Z(0)}{2 N(0)^4} \biggr) \label{eqIsoFt}
\end{equation}
as $n \rightarrow\infty$.
\end{corollary}

The proof of this corollary is analogous to the proof of Corollary~2
given in \citet{Groeneboom1995}, in our case applying Theorem
\ref{thmIsoCLT} from above. Recall that consistency at zero follows
from Lemma~1 in the supplemental article
[\citet{suppA}].

\section{Simulation}\label{secSimulation}
To validate the model and estimators, we implemented a simulation where
we work directly with the distributions of $X$ and $H$ to calculate the
distributions of $Z$ and $H$ for the rectangles, as well as the other
quantities of interest. To start, $X$ is taken to be Gamma(3)
distributed and $H$, given $X=x$, is triangularly distributed on
$[0,x]$:
%
\begin{eqnarray}\label{eq3DMarginals}
f_X(x) &=& \tfrac{1}2 x^2 e^{-x},\qquad x\geq0,
\nonumber\\[-8pt]\\[-8pt]\nonumber
f_{H|X}(h|x) &=& \frac{2}{x^2} (x-h),\qquad h\in(0,x).
\end{eqnarray}
From the above, marginal and conditional densities of the observable
quantities can be calculated:
%
\begin{eqnarray}\label{eq2DMarginals}
g_Z(z) &=& \tfrac{4}{15} \bigl(z^2+z+
\tfrac{3}4 \bigr)e^{-z}, \qquad z \geq0,\nonumber
\\
g_{H|Z}(h|z) &=& \frac{2 ({1}/2+z-h )} {
(z^2+z+{3}/4 )}1_{[0<h<z]}
\\
&&{} + \frac{2 [ ({1}/2+z-h )
I_G ({1}/2,(h-z) )
+\sqrt{h-z}e^{-(h-z)} ]} {
\sqrt{\pi} (z^2+z+{3}/4 )}
1_{[h>z]},\hspace*{-27pt}\nonumber
\end{eqnarray}
where $I_G(m,x)=\int_{t=x}^\infty t^{m-1} e^{-t} \,dt$ is the incomplete
Gamma function. From the joint densities, the underlying distributions
for the various quantities of interest ($V$, $S$ and $R$) can be
calculated. As an example, the distribution function for the volume is
as follows:
%
\begin{equation}
\qquad F_V(v) = 1 - \biggl[1 + \sqrt{\frac{v}{\pi}} -
\frac{v}{2\pi} + \frac{1}2 \biggl(\frac{v}{\pi}
\biggr)^{3/2} \biggr] e^{-\sqrt{{v/\pi}}} +\frac{v^2}{2\pi^2}
Ei \biggl(
\sqrt{\frac{v}{\pi}} \biggr), \label{eqUnderlyingVol}
\end{equation}
where $Ei(x)=\int_{u=x}^\infty e^{-u} u^{-1} \,du$ is the exponential
integral. For this simulation, we draw $n$ observations from the
marginal density $g_Z$. For each observation from~$Z$, a corresponding
height observation is drawn from the conditional density $g_{H|Z}$ to
form the 2D observations $(Z_1,H_1),\ldots,(Z_n,H_n)$ for the $n$
rectangles. From these observations it is possible to estimate the
various quantities of interest for the cylinder, beginning with the
covariance between the cylinder height and radius.

%
\begin{table}
\tabcolsep=0pt
\caption{Results of the covariance estimation for
the simulation. $n$ is the number of observed rectangles on
the cut plane. $\hat{\sigma}_{n,\sqrt{X}H}$ is the
covariance estimate given in (\protect\ref{eqestcov}).
$\hat{\nu}_g^2$ is the asymptotic variance determined
from a single simulation run based on (\protect\ref{eqnug^2}) using
the empirical means as estimates for the expectations in
the equation. The fourth column gives half the width of the
constructed 95\% confidence interval for the covariance.
The fifth column gives the empirical mean from 1000
simulation runs of the covariance estimate}\label{tabSim}
\begin{tabular*}{\tablewidth}{@{\extracolsep{\fill}}@{}lcccc@{}}
\hline
\multicolumn{5}{c}{\textbf{Covariance estimator and the asymptotic variance}}\\
\hline
$\bolds{n}$ & $\bolds{\hat{\sigma}_{n,\sqrt{X}H}}$ & $\bolds{\hat{\nu}_g^2}$
& $\bolds{1.96\sqrt{\frac{\ln n}{n}} \hat{\nu}_g}$ & $\bolds{\frac{1}{1000} \sum_{i=1}^{1000}
\hat{\sigma}_{i,n,\sqrt{X}H}}$ \\
\hline
\phantom{,000}50 & $0.424$ & $1.11$ & $0.58$ & $0.266$ \\
\phantom{,00}500 & $0.331$ & $1.12$ & $0.23$ & $0.277$ \\
\phantom{,0}5000 & $0.262$ & $1.58$ & $0.10$ & $0.277$ \\
50,000 & $0.273$ & $1.53$ & $0.04$ & $0.276$ \\
\phantom{,000}$\infty$ & $0.277$ & $1.50$ & $-$ & $0.277$ \\
\hline
\end{tabular*}
\end{table}

Table~\ref{tabSim} shows the results for the estimation of the
covariance of $\sqrt{X}$ and $H$ as calculated from the 2D
observations. The first column indicates the number of observed
rectangles on the cut plane. For $n=\infty$, the true underlying
covariance and asymptotic variance are given. For this simulation, the
underlying covariance, as calculated from (\ref{eqcovariance}) and
(\ref{eq3DMarginals}), is $0.277$, and the true underyling asymptotic
variance, as calculated from (\ref{eqnug^2}) and
(\ref{eq3DMarginals}), is $1.50$. The second column gives the
estimates of the covariance for a single simulation run. The third
column gives the estimate of the asymptotic variance for the covariance
estimator for a single simulation run. The asymptotic variance was
estimated from the empirical means for the expectations in
(\ref{eqnug^2}) and by using the following estimator for
(\ref{eqxigj3D}):
%
\begin{equation}
\hat{\xi}_g^j = \frac{1}{b_n n} \sum
_{i=1}^n H_i^j
1_{[0,b_n]}(Z_i), \label{eqEstxigj}
\end{equation}
where $b_n \sim n^{-1/3}$ is a cutoff value for approximating $z=0$ and
can be shown to have an optimal vanishing rate for the MSE of $n^{-2/3}$
(see Chapter~3 of \citet{MeThesis} for details of the MSE and the
supplemental article [\citet{suppA}] for a discussion on the
affects of the
choice of bandwidth). The fourth column gives the half-widths of
the constructed 95\% confidence interval for the covariance using
the estimators for the covariance.

%
\begin{figure}

\includegraphics{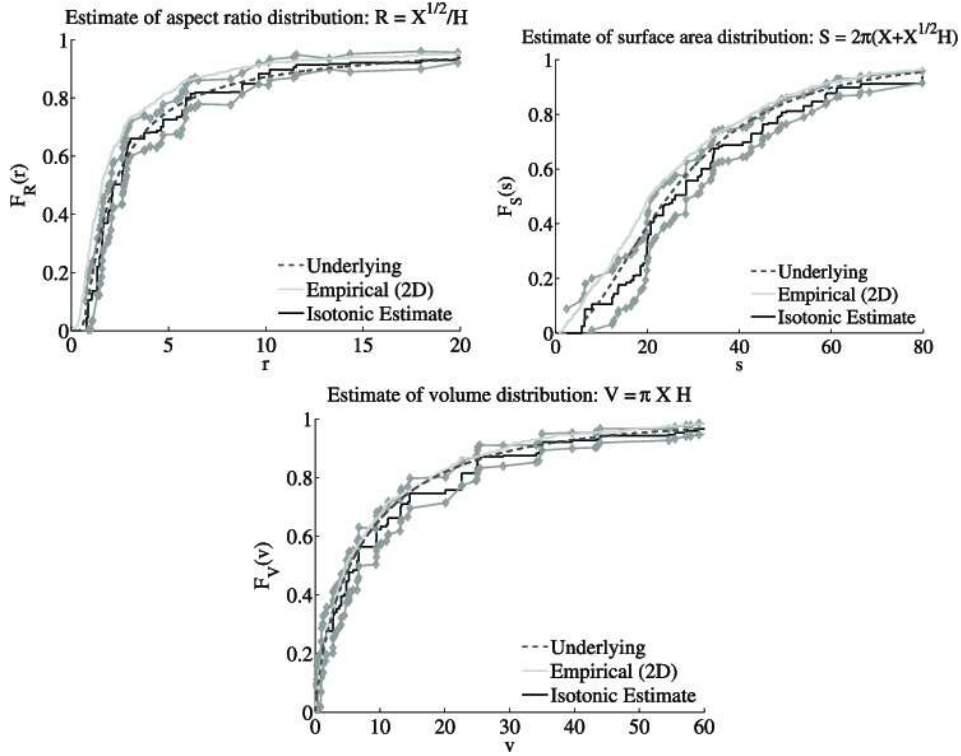}

\caption{Plots of the cumulative distribution
functions for the aspect ratio, $R$, surface area, $S$, and
volume, $V$, for $n=500$ observations of $(Z,H)$ drawn from
the 2D distributions in (\protect\ref{eq2DMarginals}). In all
figures, the dashed dark grey line gives the underlying
distribution, the light grey line gives the empirical
distribution based on the 2D observations $(Z,H)$, and the
black line gives the isotonic estimation of the
distribution of the quantity of interest based on the 2D
observations. The grey diamonds give approximate 95\%
point-wise confidence sets for the isotonic estimator.}\vspace*{-2pt}\label{figSims}
\end{figure}

The final column shows the empirical mean over 1000 simulation runs of
the covariance estimate using the 2D observations. The results behave as
expected. While the single simulation runs at small $n$ give large
values for the covariance estimate, the true covariance falls within the
constructed 95\% confidence interval. As $n$ increases, the estimated
covariance approaches the true covariance. For the mean of 1000
simulation runs, we see that the estimated value of the covariance is
much closer to the expected value, even for small $n$. This demonstrates
the consistency and unbiased nature of the estimator.

It is also possible to estimate the underlying distribution functions,
such as that given in (\ref{eqUnderlyingVol}), and compare the
empirical distribution function of the quantities of interest also based
on the 2D observations $(Z_1,H_1),\ldots,(Z_n,H_n)$ treated as if they were
distributed as the $(X,H)$. This means, for example, that for
(\ref{eqUnderlyingVol}), $V_i = \pi Z_i H_i$. This is done because
often only 2D data is accessible and has often been justified as a
reasonable approximation for the 3D data. In the case of the squared
radius, including the 2D data in this way emphasizes the bias inherent
in the observations. For the volume, surface area and aspect ratio, the
bias still exists. The question is whether the proposed estimators
provide a better estimate than using the 2D data straight up, and,
compared to the confidence intervals, it would seem that indeed it is.

Applying both the empirical and isotonic estimation procedures to the
generated data sets leads to the results for the estimation of the
aspect ratio (left), the surface area (middle) and the volume (right)
displayed in Figure~\ref{figSims}. The underlying distribution is given
by the dashed dark grey curve. The empirical distribution based on the 2D
observations [as if the $(Z,H)$ were distributed as the $(X,H)$] is
given by the light grey curve. The isotonic estimate of the distribution
of the quantity of interest based on the 2D observations is given by the
black curve. The 95\% point-wise confidence sets for the isotonic
estimator are given by the grey diamonds.

The point-wise confidence sets are calculated from the results of
Corrollary \ref{corIsoFt}. To obtain an estimate of the asymptotic
variance $\nu_g$ given in (\ref{eqIsoFt}), only the function
$\tau_q(0) = \int_{h=0}^\infty g(q(h;t),h) \,dh$ needs yet to be
estimated. The estimates for $N(0)$ and $N(t)$ can be obtained from the
isotonic estimates described in Section~\ref{secEstimation}. The
function $g_Z(0) = \xi_g^0$ and can be estimated by (\ref{eqEstxigj}).
Following the same idea as this estimator, and without going into the
asymptotic behavior, we can estimate $\tau_q(0)$ consistently as
\[
\hat{\tau}_q(0) = \frac{1}{2 b_n n} \sum
_{i=1}^n 1_{[-b_n,b_n]}\bigl(Z_i-q(H_i;t)
\bigr).
\]
The underlying distribution is mostly within the 95\% point-wise
confidence sets indicating that the estimator is reasonable and can be
used in practice.

%
\begin{figure}

\includegraphics{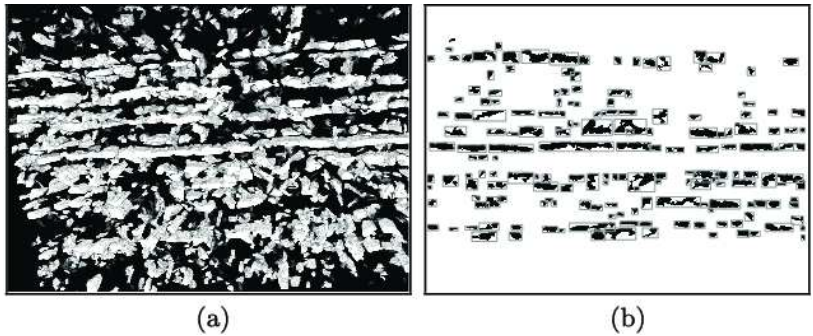}

\caption{Figure \textup{(a)} shows a
view of the 3D reconstruction of the banded microstructure
from the serial sectioned images.
Figure \textup{(b)} shows the bounding boxes around the features of interest
(heretofore referred to as rectangles and cylinders for the
2D and 3D objects, resp.) in the microstructure.}\label{fig3DView}\label{figBBs8}\label{figBoundingBoxes}
\end{figure}

\section{Application of the model to real microstructures}\label{secReal}
The model and estimation procedures are now applied to the banded steel
microstructure shown in Figure~\ref{figBandedStructure}. To obtain 3D
information about the microstructure, the material was serial sectioned,
providing images approximately every 2~\textmu m into a depth of about
90~\textmu m. For details on the experimental procedure see \citet
{Me3D}. The
optical images were processed with dilation and closing image operations
on binary thresholds. The serial sectioned images were combined to form
a single 3D object, shown in Figure~\ref{fig3DView}(a), and the bounding
boxes, that is~the smallest box that contains all voxels of the object
being considered, around the 3D features of interest (heretofore
referred to as cylinders) were found using the 3D analysis function in
Fiji [\citet{Fiji3D}]. From Figure~\ref{fig3DView}(a) it is clear
that the
3D data is incomplete. The sectioning depth was not sufficient to
observe a cylinder in its entirety. This gives a clear indication of why
using this model to estimate the distributions of the quantities of
interest is so important. Using even one of the section images, like the
one shown in Figure~\ref{figBBs8}(b), can provide a reasonable estimate
for the underlying 3D distribution that is costly to obtain directly.
Figure~\ref{figBBs8}(b) shows rectangles around the 2D features of
interest. These are the smallest rectangles to fully contain the objects
of interest and are called bounding boxes. These rectangles were found
using Fiji software [\citet{ImageJ}] and yield the observed data pairs
$(Z,H)$ used in the estimation procedures. For a discussion on using
bounding boxes to represent the rectangles and how it affects the
results of the model, see Chapter~6 of \citet{MeThesis}.

%
\begin{table}
\tabcolsep=0pt
\tablewidth=200pt
\caption{Results for the moment and covariance
estimates of the microstructure data with 179 rectangle
observations. The first column gives the estimated quantity.
The second gives the estimate of that quantity with the
half-widths of the constructed 95\% confidence intervals
using the estimates for the asymptotic variance}\label{tabReal}
\begin{tabular*}{\tablewidth}{@{\extracolsep{\fill}}@{}lc@{}}
\hline
\multicolumn{2}{c}{\textbf{Moment and covariance estimates \textbf{($\bolds{n = 179}$)}}}\\
\hline
\textbf{Quantity} & \multicolumn{1}{c}{\textbf{2D estimate $\bolds{{\pm}1.96\sqrt{\frac{\ln n}{n}} \hat{\nu}_g}$}}\\
\hline
$E [\sqrt{X} ]$ & $10.55 \pm2.30$  \textmu m\phantom{000} \\
$E[X]$ & $125 \pm27$  \textmu m$^2$\phantom{00}\\
$E[H]$ & $8.72 \pm0.56$  \textmu m\phantom{00}\\
$E[S]$ & $1434 \pm332$  \textmu m$^2$\phantom{00}\\
$E[V]$ & $4370 \pm950$  \textmu m$^3$\phantom{00}\\
$\sigma_{\sqrt{X}H}$ & \phantom{00.}$11.1 \pm25.6 < 0$  \textmu m$^2$\\
\hline
\end{tabular*}
\end{table}

Table~\ref{tabReal} gives the 2D estimates for the moments and
covariance, and the half-widths of their constructed 95\% confidence
intervals. The second column of the table gives the estimates using
(\ref{eqMomentsEst}) and (\ref{eqestcov}) for the moments and
covariance based on a single 2D estimate for the asymptotic variances
of the moments from Table~\ref{tabnu2}. For the covariance, the
estimates for the asymptotic variance come from (\ref{eqnug^2}).
%
\begin{figure}

\includegraphics{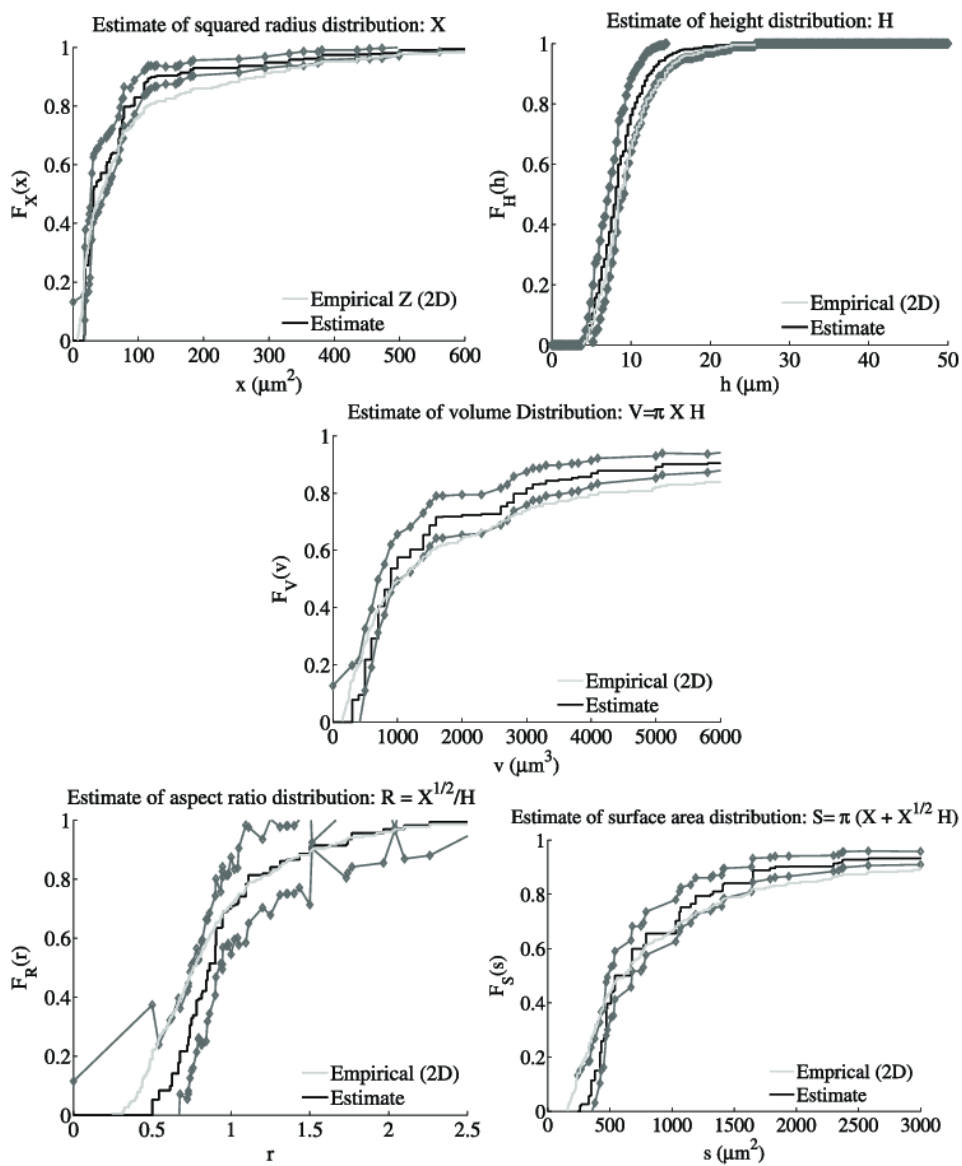}

\caption{Results of the model and estimation
procedures applied to the microstructure shown in Figure~\protect\ref
{figBoundingBoxes}. The number of observations is
$n=179$. The light grey lines are the estimates obtained by
treating the squared half-width and height of the bounding
box as if it were the squared radius and height of the
cylinder. The black lines give the isotonic estimations of
the underlying distribution functions of the quantities of
interest given the 2D observations. The grey diamonds give
the asymptotic 95\% point-wise confidence sets for the
isotonic estimates.}
\label{figRealResults}
\end{figure}

Using the 2D data set from any single slice of the serial sectioning, we
can apply the model and estimation procedures to find the CDFs of the
various quantities of interest. Figure~\ref{figRealResults} shows the
results of the estimation procedures. The upper left plot shows the
results for the isotonic estimation of the squared radius distribution.
The upper right plot shows the plug-in estimation results for the height
distribution. The middle plot and the lower left and right plots show
the results for the isotonic estimation of the distributions for the
volume, aspect ratio and surface area, respectively. In all plots, the
light grey lines show the empirical estimates obtained by treating the
rectangle squared half-width and height as if they were the squared
radius and height of the cylinder. The black lines are the isotonic
estimation results for the underlying distribution functions of the
quantities of interest given the 2D observations. The grey diamonds give
the asymptotic point-wise confidence sets for the isotonic estimates of
$F_T$ taken at the values of $t$ corresponding to the 2D observations.
These bands are calculated from the results of Corollary
\ref{corIsoFt} and Theorem \ref{thmFHh}. For the asymptotic
variance of the height distribution, the estimator $\widehat{F}_{H,n}$
is used. An estimator for the integrals again follows the same idea as
the estimator for $\xi_g^0$, and, without considering the asymptotic
behavior, we obtain
\begin{eqnarray*}
\frac{1}{b_n n} \sum_{i=1}^n
1_{[0,b_n]}(Z_i) 1_{[h,\infty]}(H_i) &
\rightarrow& \int_{y=h}^\infty g(0,y) \,dy,
\\
\frac{1}{b_n n} \sum_{i=1}^n
1_{[0,b_n]}(Z_i) 1_{[0,h]}(H_i) &
\rightarrow& \int_{y=0}^h g(0,y) \,dy.
\end{eqnarray*}

As can be seen from the plots in Figure~\ref{figRealResults}, using the
empirical 2D distributions tends to overestimate the small values and
underestimate the large values of the quantities of interest. The 2D
empirical distributions do not provide a reasonable picture for the
distribution of the 3D quantities of interest. The exception to that, of
course, is the height distribution. Due to the potential correlation
between the radius and height of the cylinders, shown by a nonzero
covariance between them in Table~\ref{tabReal}, there appears to be a
small bias in the 2D observations, leading to a slight
underrepresentation of the larger height values, yet it is still
encompassed within the point-wise confidence sets. The results of the
estimates for the covariance of the cylinder radius and height, the
estimates for the various moments and the isotonic estimates for the
CDFs of the 3D quantities of interest provide a glimpse into the
microstructure that cannot be reliably obtained from the serial
sectioned data.

\section{Discussion}
Often, it is difficult to know about the full 3D nature of the
material or
object being studied. The methods available to obtain 3D data about a
material tend to be expensive in terms of resources and time,
destructive and limited to small length scales. For instance, the total
attained depth from several weeks of serial sectioning was about 90~\textmu m for the microstructures shown in this work, while many of the
cylinders are seen to be significantly larger than that. The serial
sectioning is not enough to view a cylinder in its entirety through the
depth of the sectioning. Therefore, in industry in particular, most
information about a material is based upon 2D observations, which in
many cases is insufficient. Stereology was developed to address this
issue and to find ways to extract information about the 3D nature from
the 2D observations. However, in order to be able to do this, certain
assumptions must be made about the objects being studied. In the case of
the Oriented Cylinder Model introduced in this work, the assumptions are
that the objects in the material can be represented by circular
cylinders whose axes of symmetry are all oriented in the same direction
and that the cut through the material is along that axis. It is also
assumed that the cylinders are uniformly distributed throughout the
material. While this model is simple and the assumptions are somewhat
ideal, our observations suggest that this is a reasonable starting point
upon which more complex models can be built. The Oriented Cylinder Model
provides insight into the material that has, until now, been lacking.

Assuming the model assumptions are reasonable, estimators are used to
obtain estimates of the unknown underlying distributions of various
quantities of interest. Since so little is known about the material
studied in this work, nonparametric estimators were chosen rather than
parametric ones since not enough is known about the material to assume a
specific distribution. While parametric estimators will have a better
rate of convergence and smaller variance, the difference is of order
$\sqrt{\ln(n)}$. The flexibility afforded by the nonparametric model
makes up for this difference.

The results presented especially for the microstructure data in
Section~\ref{secReal} are informative, given how little information is
available for the 3D nature of the material. However, there are several
considerations, particularly inherent to processing the images, that have
not been considered in this particular work. Edge effects are not
accounted for in this analysis. The cylinders are considered to be
completely inbounds of the observation window. However, it is possible
that cylinders ending at the edge of the image continue beyond and this
is not accounted for in this analysis. While edge effects can be
eliminated from the simulation results presented in Section~\ref
{secSimulation}, they cannot reasonably be ignored for the
microstructure. Features of interest like microstructural bands often
deviate from perfect cylinders and are not observable as perfect
rectangles. This leads to challenges in defining the dimensions of the
observed rectangle. In this work, the bounding box around the feature of
interest was taken as the rectangle. However, using the bounding box
leads to overestimation of the heights and squared radii, though the
significance of this overestimation is not immediately known.
Determining an object of interest in an image is often done through
pixel connectivity. Even though the images have undergone morphological
processing, as described in \citet{Me2D}, it is not always
possible to
preserve the true connectivity of the objects. How this affects the
outcome of the estimation under the model assumptions is also not
immediately clear. These issues are important to consider, but are
beyond the scope of this particular work.

Despite these issues, and the simplicity of the model, the estimated
distributions for the 3D quantities of interest are practicable
representations of the underlying distributions. As a first step
toward understanding and modeling a full 3D microstructure, this work
provides a solid starting point and a reasonable approximation to what
is often not directly observable.

\begin{appendix}\label{AproofsWn}
\section*{Appendix: Relationships for the quantities of interest}
First, define the quantity of interest, squared radius, aspect ratio,
surface area or volume, as $t$. Let $(u,h)$ be the observed pair of
variables. For a fixed $h>0$ we can define $t = p(h;u)$ for each
quantity of interest. In (\ref{eqqh;t}) the inverse of $p(h;u)$ is
defined as $u = q(h;t)$. These can each be calculated as follows:
%
\begin{eqnarray}\label{eqp2q}
p(h;u) &=& \lleft\{ \matrix{ u & \mbox{(squared radius)} & t
\vspace*{3pt}\cr
\displaystyle\frac{\sqrt{u}}{h} & \mbox{(aspect ratio)} &(h t)^2
\vspace*{3pt}\cr
2\pi(u+h
\sqrt{u} ) & \mbox{(surface area)} & \displaystyle{ \biggl[\sqrt{
\frac{h^2}{4} +\frac{t}{2 \pi}}-\frac{h}{2} \biggr]^2}
\vspace*{3pt}\cr
\pi h u & \mbox{(volume)} & \displaystyle{\frac{t}{\pi h}}} \rright
\}
\nonumber\\[-18pt]\\[2pt]\nonumber
&=&  q(h;t). 
\end{eqnarray}
It is important to note for all choices of $p(h;u)$ and $q(h;t)$ that
$p(h;q(h;t)) = t$ and $q(h;p(h;u)) = u$.

The derivative of these functions with respect to the second argument
is also important. Denoting this partial derivative of $p$ with respect
to $u$ by $\dot{p}$ and the partial derivative of $q$ with respect to
$t$ by $\dot{q}$ results in
\begin{eqnarray*}\label{eqpdot}
\dot{p}(h;u) &=& \lleft\{ \matrix{ 1 & \mbox{(squared radius)} & 1
\vspace*{3pt}\cr
\displaystyle\frac{1}{2h\sqrt{u}} & \mbox{(aspect ratio)} &2h^2 t
\vspace*{3pt}\cr
\displaystyle 2\pi
\biggl(1+\frac{h}{2\sqrt{u}} \biggr) & \mbox{(surface area)} & \displaystyle\frac{1}{2\pi}
\biggl(1-\frac{h} {
2\sqrt{{h^2}/{4}+{t}/{2\pi}}} \biggr)
\vspace*{3pt}\cr
\pi h & \mbox{(volume)} &
\displaystyle\frac{1}{\pi h}} \rright\}
\\
&=& \dot{q}(h;t).
\end{eqnarray*}
%
Considering the relationship between $\dot{p}$ and $\dot{q}$ and using
the linear approximation of $q$ near $t$ yields
\begin{eqnarray*}
\dot{p}\bigl(h;q(h;t)\bigr) &=& \lim_{\varepsilon\downarrow0} \frac
{p(h;q(h;t+\varepsilon/\dot{q}(h;t)))-p(h;q(h;t))}{\varepsilon}
\\
&=&
\lim_{\varepsilon\downarrow0} \frac{t+\varepsilon/\dot
{q}(h;t)-t}{\varepsilon}
\\
&=& \frac{1}{\dot{q}(h;t)}.
\end{eqnarray*}

Finally, note that $y>q(h;t)$ if and only if $t<p(h;y)$. Recall the
expression for $W_n$ can be written in terms of the function
$\phi_{n,v}$. We can use the substitution $u=q(h;y)$ in the definition
of $\phi_{n,v}$ and obtain, for $z$ and $h$ fixed,
%
\begin{eqnarray}\label{eqchangeOfVariables}
\phi_{n,v}(z,h) &=& \int_{y=t}^{(t+\delta_n v)\wedge p(h;z)}
\bigl[z-q(h;y)\bigr]^{-{1}/2} \,dy
\nonumber\\[-8pt]\\[-8pt]\nonumber
&=&  \int_{u=q(h;t)}^{q(h;t+\delta_n
v)\wedge z}
(z-u)^{-{1}/2} \dot{p}(h;u) \,du.
\end{eqnarray}
\end{appendix}


\section*{Acknowledgments}
Thanks to our industrial partner Tata Steel Europe
for the materials and use of the research facilities
in the Metallography and Surface Analysis group.
Special thanks to Piet Kok, Koen Lammers and Karin de
Moel for discussions, guidance and valuable insights
for this project.

\begin{supplement}[id=suppA]\label{suppA}
\stitle{Supplement to ``Nonparametric inference in a stereological model with oriented
cylinders applied to dual phase~steel''\\}
\slink[doi]{10.1214/14-AOAS787SUPP} 
\sdatatype{.pdf}
\sfilename{aoas787\_supp.pdf}
\sdescription{Proofs for equation (\ref{eqE[Y2]}), Lemma~\ref{lembivarCLT}, relation (\ref{eqconvTnvec}) and Theorem
\ref{thmIsoCLT}, discussion of coverage probabilities for equation
(\ref{eqConfidenceInterval}), and discussion of equation
(\ref{eqEstxigj}).}
\end{supplement}

%

\printaddresses
\end{document}